\documentclass[preprint,showpacs,showkeys,aps,prb]{revtex4}
\usepackage[dvips]{graphicx}

\begin{document}

\title{
Nonequilibrium Time Reversibility with Maps and Walks \\
}

\author{
William Graham Hoover and Carol Griswold Hoover \\
Ruby Valley Research Institute                  \\
Highway Contract 60, Box 601                    \\
Ruby Valley, Nevada 89833 ;                     \\
Edward Ronald Smith                             \\
Faculty of Engineering                          \\
Department of Mechanical Engineering            \\
Brunel University, London                       \\
See /VOLUMES/SSD2/SMITH/BAKER.f                  \\
Prepared for the 6 September 2021 Conference
on Nonequilibrium Molecular Dynamics
}
\date{\today}


\vspace{0.1cm}

\begin{abstract}
Time-reversible dynamical simulations of nonequilibrium systems exemplify both
Loschmidt's and Zerm\'elo's paradoxes. That is, computational
time-reversible simulations invariably produce solutions consistent
with the {\it irreversible} Second Law of Thermodynamics (Loschmidt's) as
well as {\it periodic} in the time (Zerm\'elo's, illustrating Poincar\'e
recurrence). Understanding these paradoxical aspects of time-reversible
systems is enhanced here by studying the simplest pair of such model systems.
The first is time-reversible, but nevertheless dissipative and periodic, the
piecewise-linear compressible Baker Map.  The fractal properties of that
two-dimensional map are mirrored by an even simpler example, the one-dimensional
random walk, confined to the unit interval. As a further puzzle the two models yield
ambiguities in determining the fractals' information dimensions. These puzzles,
including the classical paradoxes, are reviewed and explored here. We review
our investigations presented in Budapest in 1997 and end with presentday questions posed
as the Snook Prize Problems in 2020 and 2021.

\end{abstract}

\maketitle

\section{Introduction}

The averages produced by molecular dynamics and Gibbs' statistical mechanics
agree for ``ergodic'' systems, systems with a dynamics able to access all of
the $(q,p)$ coordinate-momentum states included in Gibbs'  statistical averages.
Ergodicity is actually a purely theoretical construct for manybody systems.
The time required for a nearly complete averaging over a many-dimensional
phase space grows exponentially with system size, and exceeds the age of the
universe when the number of degrees of freedom is a dozen or so. On the other hand
the ergodicity
of few-body models can be established and studied. Two hard disks, with periodic
boundary conditions is a simple example. The eight-dimensional phase space can
be reduced to three (enough for chaos) by imposing symmetry and constant energy
on the dynamics. A further simplification can be attained by considering maps,
in which the ``next'' system state is a function of the ``current'' state of the
system. The two simplest such maps are described here, the two-dimensional Baker
Map and a closely-related one-dimensional Confined Random Walk.

\subsection{Hopf's Equilibrium Baker Map E(x,y)}

Nearly a century ago Hopf introduced his Baker Map, reminiscent of a bread baker's
dough-kneading action. We construct it here in the unit square, $0<x,y<1$. See
{\bf Figure 1}. The mapping at the top is based on two choices for subsequent points:
For a current (green) point with $x<1/2$ we choose
$$
x_{\rm  new} = 2x \ {\rm and}  \ y_{\rm new} = (y+1)/2 \ , 
$$
while for a current (red) point with  $x>1/2$ we choose instead
$$
x_{\rm  new} = 2x-1 \ {\rm and} \ y_{\rm new} = y/2 \ .
$$
Hopf's interest was ``ergodic theory'' and this Baker model can be proved ergodic
{\it most} of the time.

Here ``most'' of the time means choosing an irrational initial
condition. Rational beginnings or numerical finite-precision simulations lead promptly
to fixed points instead. The fixed-point mechanism is simply the repeated doubling of the
fractional part of the $x$ coordinate. In fact {\bf Figure 1} shows the typical fate of a
numerical implementation of Hopf's equilibrium map. The quadruple-precision simulation
illustrated in the figure generates 112 $(x,y)$ states, of which the last is a fixed point.
Single and double-precision simulations likewise come to fixed-point ends, after 23 and 52
iterations of Hopf's map. If Hopf's evolution is instead described in terms of a rotated
$(q,p)$ coordinate system the result is a relatively long periodic orbit rather than a
fixed point. Even single precision gives a period of hundreds of thousands of iterations.
See again {\bf Figure 1}. 

In the summer of 1997 at the E\"otv\"os University
school/workshop meeting  ``Chaos and Irreversibility'' Hoover and Posch\cite{b1} introduced
time-reversibility and dissipation into a generalized Baker Map. Tasaki, Gilbert, and
Dorfman\cite{b2} analyzed families of similar maps at that same meeting. See as well George
Kumi\^c\'ak's related work\cite{b3} from 2004. Next, we focus on the specific
dissipative and time-reversible $(q,p)$ map considered by Hoover and Posch, and described in what
follows.

\begin{figure}
\includegraphics[width=3.5in,angle=-90.]{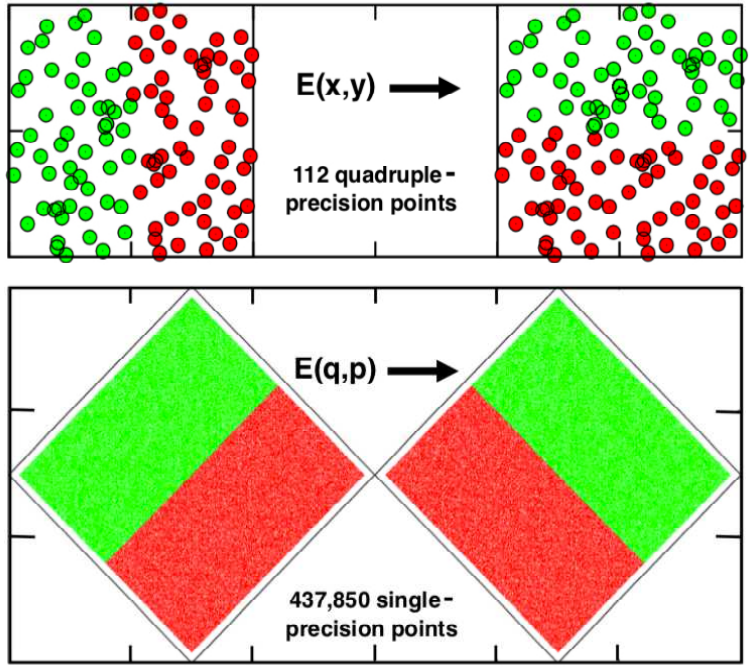}
\caption{
Hopf's deterministic Baker Map E$(x,y)$ maps the left/right sides of the unit square
into the top/bottom halves with each iteration. The continued doubling in the $x$
direction eventually reaches a fixed point. The Figure shows a series of 112 points
generated with quadruple-precision arithmetic. Evidently Hopf could not imagine
a computational implementation of his map! The diamond-shaped lower version of
the map, $2 \times 2$ rather than a unit square, produces a long periodic orbit
and is, unlike Hopf's, time-reversible. 
}
\end{figure}

\subsection{From Equilibrium to Nonequilibrium with Time-Reversible Maps}

Molecular dynamics is ``time-reversible'' if the previous step can be recovered by a three-step
process: [1] At time $t+dt$ change the signs of the momenta, $\{ \ +p \rightarrow -p \ \}$; [2]
Propagate the resulting reversed $(q,-p)$ state (backward) to time $t$; [3] Change the signs of
the momenta, matching the original $(q,+p)$ state at time $t$. A time-reversible Baker Map
B$(q,p)$ would obey the relation B$^{-1} =$ T$ \times B \times$ T, where T is the (time
reversal) mapping that reverses the momenta. If we choose $(q,p)$ coordinate-momenta variables
in a $2 \times 2$ rotated Baker map, the ``equilibrium'' (incompressible) map E, illustrated in
{\bf Figure 1} has the analytic form
\begin{verbatim}
if(q < p) qnew = (5/4)q - (3/4)p + 3d ; pnew = -(3/4)q + (5/4)p - d
if(q > p) qnew = (5/4)q - (3/4)p - 3d ; pnew = -(3/4)q + (5/4)p + d\end{verbatim}
where {\tt d = sqrt(1/8)}.
This ``motion" is analogous to ordinary Hamiltonian mechanics, where the phase volume $dqdp$
is conserved by Hamilton's motion equations. In {\it nonequilibrium} molecular dynamics the
extraction of heat, corresponding to entropy loss, leads to a continuous loss of phase
volume. A mapping analogy can be illustrated by constructing a compressible mapping, as
shown in {\bf Figures 2 and 3}.

\begin{figure}                                                                                    
\includegraphics[width=2.5in,angle=0.]{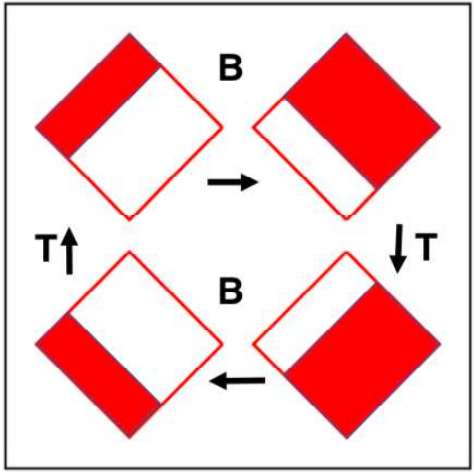}                                                  
\caption{                                                                                         
The deterministic Baker Map B doubles an area $dqdp$ in the red region and halves                 
an area in the white. The time-reversal Map T changes the sign of the vertical                    
``momentum-like'' variable p. The diamond-shaped domain of the map is                             
$| \ q \pm  p \ | < \sqrt{2}$. A counterclockwise circuit of the four states follows if B         
is replaced by B$^{-1}$ as T and T$^{-1}$ are identical.                                          
}                                                                                                 
\end{figure}                                                                                      
                                                                                                  
\begin{figure}                                                                                    
\includegraphics[width=2.5 in,angle=-90.]{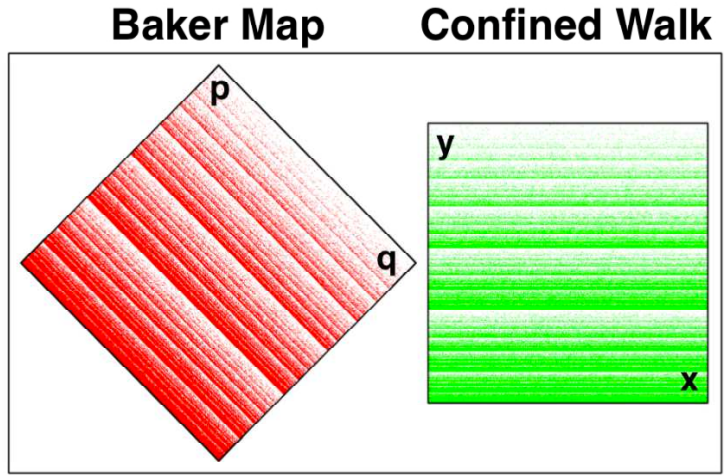}                                               
\caption{                                                                                         
One million iterations of the Baker Map and the Confined Walk are compared.                       
A scaling and translation                                                                         
of the Baker Map solution at the left to the unit square replicates the solution of a stochastic  
confined random walk problem where $x$ and $y$ are stochastic variables. The walk confined        
to the unit interval $0<y<1$ is generated with a random number relating the ``next''              
value of {\tt y} to the ``last''. The latter is either {\tt y/3} or {\tt (1+2y)/3},               
corresponding to steps to the bottom third or upper two thirds of the green million-iteration
solution at the right of {\bf Figure 3}. The $(q,p)$ Baker Map at the left and the $(x,y)$
Confined Walk at the right provide indistinguishable fractals when rotated 45 degrees and
scaled by a factor of two, as shown in the figure. The Confined Walk shown there occupies a unit
square, $0<(x,y)<1$. We show a $2\times 2$ version, with $|q \pm p|<\sqrt{2} \leftrightarrow
|x,y|<1$, here to clarify the details of the fractal structure..                                         
}                                                                                                 
\end{figure}

Nonequilibrium Molecular Dynamics has a half-century history of providing
simulations of viscous and heat-conducting flows driven by boundary differences
in velocity and temperature respectively\cite{b4,b5}. Stationary state
simulations are typically stabilized by time-reversible thermostat forces, linear
in momentum. These ``thermostat forces'' impose the desired thermalized boundary
conditions\cite{b6,b7,b8} to maintain nonequilibrium steady states. Such
simulations are irreversible despite their time-reversible motion equations. These
simulations invariably provide the positive viscosities and heat conductivities
associated with Loschmidt's paradox.

Phase-space analyses of small-scale steady-state nonequilibrium simulations indicate stable
periodic steady-state structures, ``attractors'' in phase space. The attraction is
termed ``strange'' because constrictive attractor dynamics simultaneously exhibits expansive
Lyapunov instability, with small perturbations growing exponentially in time. Despite this
expansive instability the phase volume comoving with a trajectory point is attractive,
shrinking with time as the computation settles onto a periodic, but Lyapunov unstable, orbit.
This periodicity illustrates Zerm\'elo's recurrence paradox and implies that the actual
dimensionality of computational nonequilibrium steady states is only unity. But, as the
computational precision is further refined the orbit lengthens, with the length soon
becoming too long to measure and with the one-dimensional trajectory defining a natural
measure (or coarse-grained probability density) with a fractal information dimension in
the phase space.

Thus precise long-time nonequilibrium trajectories in phase space come to define
{\it fractal} structures, still space-filling, but only sparsely. The dimensionality of
these structures is significantly less than that of the equilibrium phase space supporting
their nonequilibrium dynamics. The ``information dimensions'' of these fractals can be
estimated by phase-space binning, as we shall demonstrate. In fact such phase-space
dimensionality descriptions are not completely clearcut as their description with simple
models like Baker's reveals a sobering nonuniform convergence. The simplest model flows
exhibiting these interesting fractal formations are one-particle systems with
three-dimensional phase spaces\cite{b9,b10}. Those spaces include a  single
coordinate-momentum pair along with a time-reversible friction coefficient $\zeta$. The
friction coefficients stabilize nonequilibrium steady states.

In the remainder of this work we discuss the two simplest models, a two-dimensional
Baker Map and a related one-dimensional confined random walk.  Their study sheds light
on the coexistence of time reversibility with the dissipation typifying strange
attractor structures in model phase spaces.

\section{The Time-Reversible Dissipative Baker Map}

Maps, as opposed to flows, can exhibit similar time-reversible dissipation while
occupying only one or two phase-space dimensions.  We consider here a compressible version
of the ``Baker Map''\cite{b1,b2,b3}. It evolves a single coordinate-momentum pair of
variables $(q,p)$ as is illustrated in {\bf Figure 2}. A longtime solution appears at the
left in {\bf Figure 3}. In the two-dimensional $(q,p)$ coordinate-momentum
phase space compressible Baker-Map dynamics simply generates a new $(q,p)$ pair from the old one,
as described by a linear map. At the top of {\bf Figure 2} the smaller red area with
$q-p < \sqrt{(2/9)}$ is expanded twofold by the Baker Map B. The expanding map has the
analytic form:
\begin{verbatim}
qnew = +(11q/6) - (7p/6) + 14d ; pnew = -(7q/6) + (11p/6) - 10d .
\end{verbatim}
The constant {\tt d} is $\sqrt{(1/72)}$. Notice that the expanding map has a $(q,p)$
Jacobian determinant of $(121-49)/36 = 2$, signalling a doubling of area
with each iteration. In the larger white region the map, likewise linear, contracts:
\begin{verbatim}
qnew = +(11q/12) - (7p/12) - 7d ; pnew = -(7q/12) + (11p/12) - d .
\end{verbatim}
Here the determinant of the contracting map is $(121-49)/144 = 1/2$ signifying twofold compression.
{\bf Figure 3} illustrates a million-iteration solution of the mapping equations.

{\bf Figure 2} illustrates the time-reversibility of the map.  First, starting at the
upper right of the figure, change the sign of the vertical ``momentum variable'' $p$
with the map ``T'', ending at lower right; next map forward with ``B'' to lower left;
last reverse time again with ``T'' returning $(q,p)$ to its original top left pre-mapped
location, demonstrating that the inverse mapping is B$^{-1}$=TBT. This identity defines
a time-reversible mapping.

Though the map looks even-handed compression invariably wins out over
expansion. It must! A little reflection identifies compression with stability and
expansion with its opposite, instability. An unphysical hypothetical system in
which expansion wins out over compression would correspond to numerical
instability with an exponential divergence of the comoving phase volume. In
numerical work only about ten percent of the simple Baker-Map iterations are
computationally reversible in the sense that applying the inverse mapping undoes
the most recent iteration precisely so that B$^{-1}$B leaves $(q,p)$ unchanged.

\section{Irreversibility Through Shrinking Phase Volume}

\begin{figure}                                                                                           
\includegraphics[width=2. in,angle=0.]{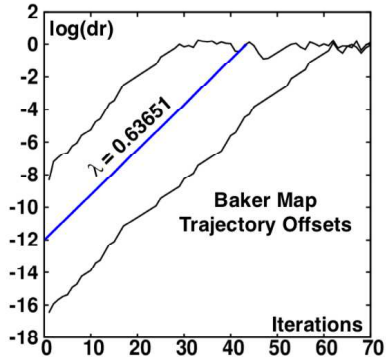}                                                         
\caption{                                                                                                
Comparison of the differences $dr = \sqrt{q^2 + p^2}$ between single and double                          
precision simulations (above) of the Baker Map and double and quadruple precision                        
simulations (below) with all three trajectories started at the origin. The straight                      
blue line has a slope corresponding to the largest Lyapunov exponent,                                    
$\lambda_1 = 0.63651$.                                                                                   
}                                                                                                        
\end{figure}  

An understanding of the shrinking phase volume, which leads to an apparent
fractal (fractional dimensional) phase-space object is straightforward in
the Baker Map example of {\bf Figure 2}. The map, whether in the red region
or the white, always expands in the northwest-southeast direction parallel
to lines of constant $q+p$. This expansion characterizes Lyapunov instability,
the growth of microscopic perturbations to macroscopic scale. In the
perpendicular direction, parallel to lines of constant $q-p$, the map
contracts to the self-similar fractal structure displayed in {\bf Figure 3}.

\subsection{Lyapunov Instability and Exponential Growth}

{\bf Figure 4} illustrates the exponential character of the expansive
northwest-southeast  growth
by displaying the offset between single- and double-precision simulations, the
uppermost of the three curves in the figure. Both simulation types begin at the
$(q,p)=(0,0)$ origin at the center of the diamonds shown in {\bf Figures 1
through 3}. The initial separation immediately reflects the roundoff error of the
single-precision mapping, of order $10^{-10}$. The single-double separation
increases by ten orders of magnitude in about 40 iterations of the map.
Similarly, the double-quadruple separation likewise grows exponentially. In
that more-nearly-accurate more-precise case the growth rate is the same
$~ e^{+\lambda_1t}$. $\lambda_1$ is the largest Lyapunov exponent. Its
numerical value is 0.63651. The exponent and the blue middle line in
{\bf Figure 4} drawn with its slope corresponds to averaging the growth rates
in the red and white regions, $\ln(3)$ and $\ln(3/2)$ respectively, taking
into account that the white compressive region is twice as likely as the
red. The result is the time-averaged expansion rate:
$$
\langle \ \lambda_1 \ \rangle = (1/3)\ln(3) + (2/3)\ln(3/2) =
(1/3)\ln(27/4) = 0.63651 \ .
$$
Similarly, the compression perpendicular to the expansion gives the
second Lyapunov exponent:
$$
\langle \ \lambda_2 \ \rangle = (1/3)\ln(2/3) + (2/3)\ln(1/3) = 
(1/3)\ln(2/27) = -0.86756 \ . 
$$
This analysis is in good agreement with the numerical data. With an initial
double-precision roundoff error of order $10^{-17}$ the exponential loss of
accuracy expands to unity after about 60 iterations of the map:
$10^{+17} \simeq e^{0.63651\times 60}$, as in the lower curve.

\subsection{Compression with Expansion Leads to Irreversibility}

\begin{figure}                                                                                                
\includegraphics[width=2. in,angle=0.]{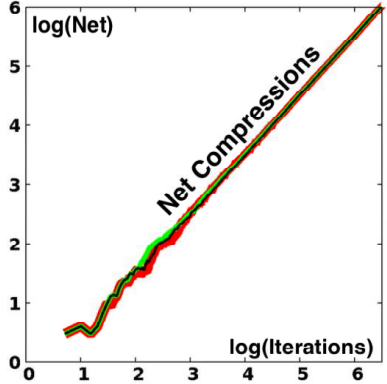}                                                              
\caption{                                                                                                     
Excess of compressive over expansive iterations of the single-, double-,                                        
and quadruple-precision Baker maps. The differences between them are visible                                        
between 20 and 1000 iterations of the maps. The final values of the                                           
excess are 1000742, 1000250, and 998236 for the three sets of three                                           
million iterations beginning at $(0,0)$.                                                                      
}                                                                                                             
\end{figure} 

Despite the exponential growth of northwest-southeast separations the
overall phase volume shrinks. On average a white area is halved two thirds
of the time while a red area doubles one third of the time.
Thus overall the comoving area decreases as $2^{(-t/3)}$ with $t$ iterations.
That area is soon reduced to a vanishingly-small fraction of its initial value.
That fraction is of order $e^{-100}$ for a thousand iterations of the map.
This contraction of area accounts for the sparse appearance of the thousands
of mapped points in the smallest-scaled regions of {\bf Figure 3}. The mean
densities in the self-similar bands decrease with increasing values of $y$ :
$$
{\textstyle
\langle \ \rho(0<y<\frac{1}{3}) \ \rangle = 2^1 \ ;
\langle \ \rho(\frac{3}{9}<y<\frac{5}{9}) \ \rangle = 2^0 \ ; 
\langle \ \rho(\frac{15}{27}<y<\frac{19}{27}) \ \rangle = 2^{-1} \dots \ .
}
$$
The expansive ``strange'' portion of the mapping is responsible for this
decrease in density even though it is unlikely relative to compression. Consider
an initial point at the origin $(q,p) = (0,0)$ and follow it forward in time,
counting the net number of compression steps. {\bf Figure 5} shows the results of
three million iterations of the map in single,  double, and quadruple precision.
As expected the net numbers of compressive iterations are ``close to'' [within a
thousand or so] one-third the total.

\section{Poincar\'e Recurrence of the Baker Maps}

A fringe benefit of the diamond-shaped Baker Map is the relatively long
Poincar\'e recurrence time. The computational ``noise''  contributing to this
longevity can be traced to the irrational square roots in the mapping. The
single-precision mapping which we have used in {\bf Figures 4 and 5} has a periodic
orbit of length just over a million iterations, 1,042,249 to be precise.  The
double-precision mapping settles into a periodic orbit repeating after a few
trillion iterations\cite{b11}. The $(x,y)$ Cartesian coordinate version of the
map corresponding to the orientation at the right in {\bf Figure 3} is not well
suited to computation due to its many short periodic orbits, many of which are
stable (which we view as ``unphysical''. A clear and comprehensive analysis of
the generalized Baker Map problem (with arbitrarily small and large expansions
and contractions) was presented by  Kumi\^c\'ak in 2005\cite{b3}.

\section{Characterizing Chaos in the Baker Map}

\begin{figure}                                                                                                     
\includegraphics[width=1.5 in,angle=-90.]{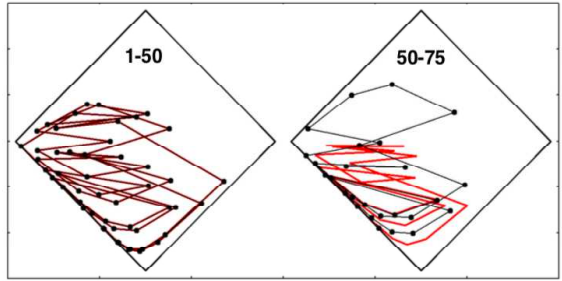}                                                                
\caption{                                                                                                          
50 (at the left) and 25 additional (at the right) iterations in double (black)                                     
and quadruple (red)                                                                                                
precision for the two-dimensional $(q,p)$ Baker Map B with the initial point                                       
at the origin $(0,0)$. Lyapunov instability makes the difference between the                                       
two solutions visible after about 70 iterations, as can be seen at the right.                                      
}                                                                                                                  
\end{figure}  

Numerical solutions carry a fixed number of digits, 9, 17, and 36
for single-, double-, and quadruple-precision numbers on the unit
interval. Chaos is characterized by the exponential growth of small
perturbations. The Baker Map's  largest Lyapunov exponent is 0.63651,
so that the differences seeded by roundoff between single- and
double-precision, and between double- and quadruple-precision
solutions can be estimated from {\bf Figure 4} just as easily as
from the analytic growth rates. That figure  illustrates the offsets and
{\bf Figure 6} the trajectory differences between 50 and 75 iterations of
the double and quadruple-precision maps. At 50 iterations,
corresponding to multiplying by $e^{32} \simeq 10^{14}$, roundoff error
hasn't yet amplified the difference between double and quadruple 
precision to visibility, while 75 iterations are more than
sufficient to lose any visible correlation between the two solutions.

The right side of {\bf Figure 3} shows that a stochastic view of the map, traceable
to its chaos, is quite proper. Before the mapping shown at the top
of {\bf Figure 1} is executed we note that the white southeastern
area is twice that of the northwestern red area so that compressive
steps (into the highest-density lower third of the unit square)
are twice as likely as expansive steps (into the upper two-thirds).
The preponderance of the excess, lower - upper, is
shown in {\bf Figure 5} and approaches, on average, $t/3$ as the number
of iterations $t$ increases. We would expect the error in
this statistical estimate to become visually negligible, say one
percent, once the number of iterations is of order $10^4$. A look
at the figure shows that the statistical bumpiness away from a straight line
becomes visually negligible between 1000 and 10,000 iterations, just as one
would expect for a stochastic, rather than deterministic, process.

It is an article of faith that the $x$ motion of the $(x,y)$ map
is completely random\cite{b12}. This statistical view is consistent with our
numerical work so that we believe it is fully justified.
It is easily checked by comparing bin populations for the Map and the
Walk after millions or billions or trillions of iterations.  See
{\bf Figure 7} for a million-iteration sampling of 2187 bins for
both approaches. Let us consider the confined random walk problem
in more detail next\cite{b11,b12,b13}.

\begin{figure}                                                                                                           
\includegraphics[width=1.5 in,angle=-90.]{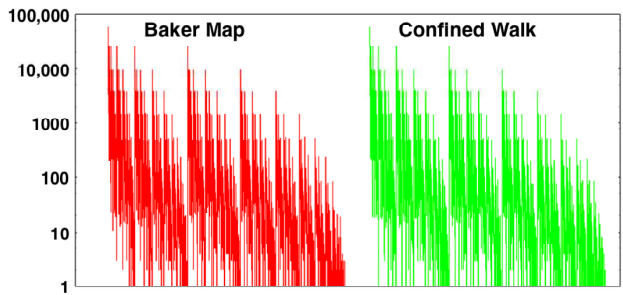}                                                                      
\caption{                                                                                                                
Comparison of the $y$ coordinate distribution of the Baker Map in the                                                    
unit square converted from $(q,p)$ with the Confined Walk distribution                                                   
obtained using the FORTRAN random-number generator {\tt random}$_-${\tt number(r)}.                                      
The Map and Walk data, one million points for each, have been collected here and                                         
displayed in $2187 = 3^7$ bins of width $3^{-7}$.                                                                        
}                                                                                                                        
\end{figure}  

\section{Random Walk Analog of the Baker Map}

A random-walk analog for the progress of the Cartesian $y$
variable on the unit interval can be modelled given current values
of {\tt x} and {\tt y}. A new value {\tt xnew} can be chosen at
random, while {\tt ynew} depends upon both a random number {\tt r}
and the current value of {\tt y}:
\begin{verbatim}
if(r.lt.1/3) ynew = (1+2y)/3 or if(r.gt.1/3) ynew = y/3
\end{verbatim}
The righthand side of {\bf Figure 3} illustrates the strange-attractor
character of numerical solutions of this confined walk problem. The
Confined Walk fractal at the right is related to the Baker Map fractal at the
left by a 45-degree counterlockwise rotation. Evidently, due to the
exponential growth in the walk's horizontal $x$ direction that
distribution is random and can be modelled by a good
random-number generator. In the $y$ direction the distribution is
a self-similar fractal, repeating in an infinite set of bands with
each band smaller than its predecessor by a factor of (2/3) and
including one-third as many points.

{\bf Figures 3 and 7} compare the distributions from one million iterations
of the two-dimensional Baker Map with those from the same number of iterations
applied to the confined walk. In {\bf Figure 7} the unit-interval $y$ values
have been ``binned'' into $3^7 = 2187$ bins of equal width, $\delta = 1/2187$.
Initially the doubling of density with steps to the bottom third of the square
and halving with steps to the upper two thirds gives rise to probability
density steps of factors of four. Later, in the steady state and visible in the
plot where the bin probabilities are plotted on a logarithmic scale spanning 16
e-foldings, the regular steps stand out in the lefthand 3-based binnings but are
less distinct in the righthand 4-based ones. This difference, along with the
continual increase in e-fioldings with iterations, suggests the possibility of
convergence difficulties in characterizing the resultant fractal.

The only singularity in the linear Baker Map is the border line
separating the red and white regions:
$$
q-p = -\sqrt{(2/9)} \ {\rm Baker \  Map} \ \longleftrightarrow
 x = (1/3) \ {\rm Random \ Walk}
$$
Evidently this measure-zero set of singular points is enough to generate the
everywhere-singular fractal distribution of the analytic ``$\{ \ y \ \}$''
and the computational ``$\{ \ {\tt y} \ \}$''. Let us consider further details
of the latter set next.

\section{Nonuniform Convergence of the Information Dimensions}

\begin{figure}                                                                                                             
\includegraphics[width=2.5in,angle=-90.]{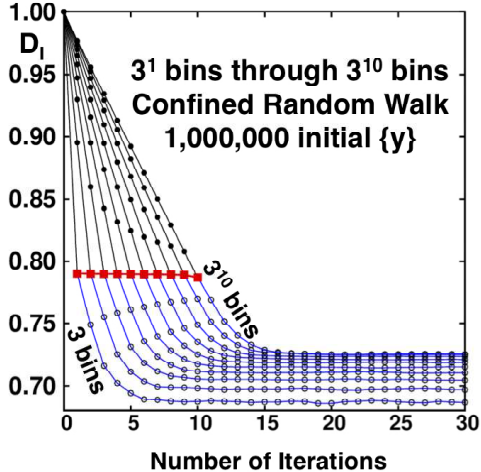}                                                                         
\caption{                                                                                                                  
The information dimensions $D_I$ of the developing random-walk fractal as ten functions                                    
of the number of iterations. The number of bins characterizing each curve varies from                                      
$3^1$ to $3^{10}$ for each of thirty iterations. Each point corresponds to an averaged                                     
$D_I$ from one million equally-spaced initial conditions on the unit interval $0<y<1$.                                     
In the limiting special cases (shown as red squares) that the number of iterations is                                      
equal to the logarithm, base-3, of the number of bins, the information dimension follows                                   
from Farmer's analysis\cite{b11},  $[(2/3)\ln(2/3) + (1/3)\ln(1/6)]/\ln(1/3) = 0.78967$.                                   
When the number of iterations approaches infinity ahead of the number of bins, the                                         
dimensionality is substantially lower, $D_I \simeq 0.741_5$ rather than the Kaplan-Yorke                                   
conjectured value (based on the Baker-Map Lyapunov exponents) 0.7337.                                                      
}                                                                                                                          
\end{figure}

The definition of the information dimension $D_I(\delta)$ for a set of points
describes the dependendence of the probability density of points on the size of the
sampling bins, $\delta$. Although other fractal dimensions can and have been defined
and studied, the information dimension is uniquely significant.  Unlike the correlation
dimension $D_I$ is unchanged by simple coordinate transformations\cite{b14}.

To seek uniqueness both the number of bins and the number of
points per bin must approach infinity in averaging the bin probabilities:
$D_I = \langle \ \ln({\rm prob}) \ \rangle /\ln(\delta)$. To visualize taking this
limit we illustrate the result of iterating a set of one million equally-spaced initial
points on confined random walks: $0 < \{ y_i(t) \} < 1$. We explore thirty iterations:
$0<t<31$. The confined random-walk iterations are governed by the output of the FORTRAN
random number
generator {\tt random}$_-${\tt number(r)}. The dimensionality data are analyzed here
using $3^n$ bins, with  $n$ varying from 0 to 10. The finest grid has $3^{10} = 59,049$
bins of equal width $\delta = 1/59049$.  By combining the contents of 3, or 9, or 27,
or ... contiguous bins the entire set of 30 stepwise information dimensions for the ten
binnings choices can be obtained from a single run.  The apparent information dimensions
for the 300 problems (thirty iterations with ten bin sizes) are plotted as the ten lines
shown in {\bf Figure 8}.

The Baker-Map function, $y=[(q+p)/\sqrt{2} + 1]/2$ provides the same fractal as does the
confined walk, penetrating, in both cases, to a scale smaller by a factor 3 with each
iteration. For this reason powers of $(1/3)$ are the ``natural'' bin sizes for analyzing
the Baker-Map function\cite{b11,b12} and the confined walk $0<y<1$.
Although reciprocal bin widths which are powers of 3 are ``natural'' for the Baker Map
and its confined walk analog, an embarrassing variety of choices is possible. As an example
bin widths which are the first eight powers of 4 (a subset of bin widths which are powers
of 2) provide the
information dimension estimates shown in {\bf Figure 9}. For additional examples see
Reference 14. The totality of these results is paradoxical because they indicate
a limiting information dimension of $0.741_5$ from the series of widths $(1/3)^n$ and a
{\it different} limit, $0.7337$, from the series of widths $(1/4)^n$. This difference
suggests a persistent difference of distributions in the limiting case(s) $\delta
\rightarrow 0$. This nonuniform convergence caught us completely by surprise.

\begin{figure}                                                                                                                 
\includegraphics[width=2.5in,angle=-90.]{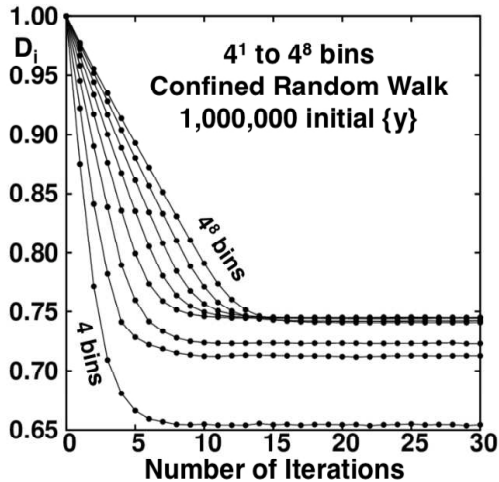}                                                                             
\caption{                                                                                                                      
The information dimension $D_I$ of the developing random-walk fractal as a function of                                         
the numbers of bins and iterations. The number of bins varies from $4^1$ to $4^{8}$ for                                        
thirty iterations. As in {\bf Figure 7} each point corresponds to $D_I$ for one of the ten                                     
samples of one million initial conditions.                                                                                     
}                                                                                                                              
\end{figure}   

The dependence of the limiting information dimension on the bin-width power law, giving
either $0.741_5$ or $0.7337$, suggests a look at the distributions themselves. As the
limiting case(s) are singular everywhere, we arbitrarily choose to compare probability
densities for both $3^{10}$ and $4^8$ bins in {\bf Figure 10}. Both simulations include
exactly the same set of $100,000,000$ iterations. The density steps with $3^{10}$ bins
are markedly sharper than those with $4^8$ and the details of the boundaries between
vertical strips are likewise better defined for the finer (65536 rather than 59049 bins)
mesh.

\begin{figure}                                                                                                                    
\includegraphics[width=2.5in,angle=-90.]{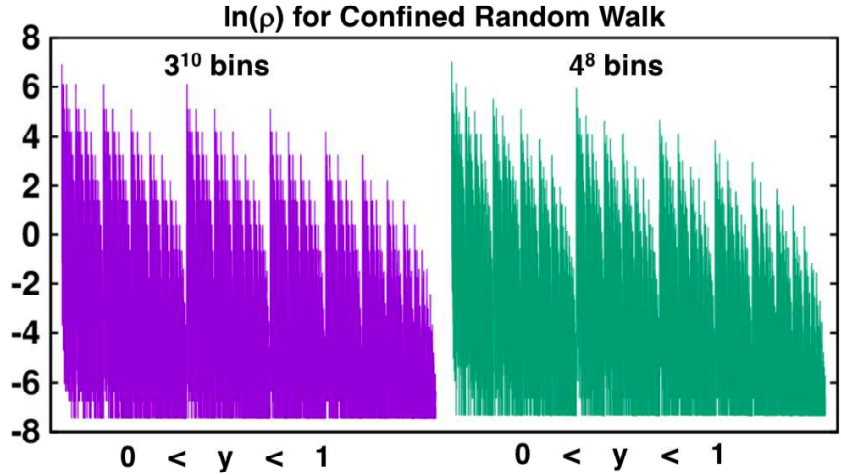}                                                                                
\caption{                                                                                                                         
Fractal probability densities for confined walks with bin sizes $\delta = 3^{-10}$ and                                            
$4^{-8}$. For both bin sizes $\int_0^1\rho(y)dy = 1.$ 100 million iterations of the map were used for                             
the point set that was analyzed with both these choices of binning.                                                               
}                                                                                                                                 
\end{figure}  

A clearer picture of the binning dependence follows from the cumulative distributions
of density and information, shown for the same data used in {\bf Figure 10}. {\bf Figure 11}
shows both the density and the information dimension as cumulative sums for two pairs of
similar binnings, $(1/3)^5 \sim (1/2)^8$ and $(1/3)^7 \sim (1/2)^{11}$. Because the underlying
data are identical the densities always agree, within one bin width. The information 
dimensions are quite different with powers of 2 both giving $D_I=0.745$, significantly
great than $D_I = 0.715$ and 0.725 for the ``natural choice" of powers of three, 243 and
2187 bins. Numerical work suggests that the difference persists even to infinitesimal bin widths.

\begin{figure}                                                                                                                        
\includegraphics[width=2.0in,angle=-90.]{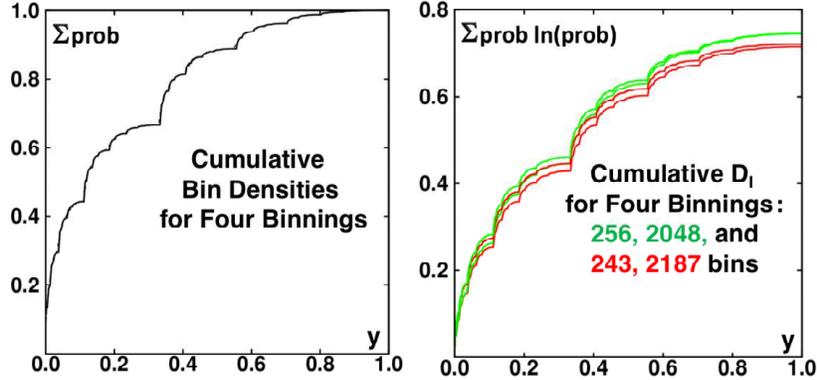}                                                                                   
\caption{
Cumulative densities and information dimensions are compared for bin widths (from bottom
to top at the right) of $(1/3)^5, \ (1/3)^7, \ (1/2)^{11}, \ (1/2)^{8}$. The information
dimensions for much narrower bins suggest different limiting dimensionalities for the two
bin families.
}                                                                                                                                     
\end{figure}  

\section{Summary and the 2021 Snook Prizes}

The finding that nonequilibrium steady states with time-reversible motion
equations generate repellor-attractor pairs in phase space has been
explored here for two simple models, the Confined Walk and the Baker Map.
These models, with one- and two-dimensional phase spaces, are simpler
than the continuous flows illustrating chaos\cite{b9,b10}.  The two model systems
reveal the details of the singular loss of phase volume resulting in
these interesting fractal distributions. Flows, with differential equations
rather than maps, require three phase-space dimensions for chaos.

The maps have shown us that nonequilibrium states are rare, occupying only fractional dimensional
portions of phase space. Although the time-symmetry of the motion equations guarantees that there
is a mirror-image $p \rightarrow - p$ solution of the motion equations our investigation shows that the
dissipative states form an attractor, with a negative Lyapunov-exponent sum. The attractor is a
fractional-dimensional stable sink relative to the two-dimensional equilibrium phase space. The
paradoxical time-reversed repellor states have positive Lyapunov-exponent sums and are wholly unstable
and irreversible.

The structures of nonequilibrium steady-state phase space flows are qualitatively different
to those of Gibbs' equilibrium ensembles. The steady state
flows are directed {\it  from} repellors {\it to} attractors. The barrier to reversal is exponential.
Loschmidt's reversed states are nearly unobservable, like the attractor states so rare that
they never turn up for long. Zerm\'elo's recurring states are simply typical non-paradoxical
dissipative states on stable attractors occupying a fractional-dimensional zero-volume portions
of their phase spaces.
Thus nonequilibium systems are qualitatively different to Gibbs' space-filling distributions of points. It
is interesting to note that the nonequilibrium fractals have the form of periodic orbits which
cannot be reversed. Even for the simple Baker Map only about ten percent of the timesteps can be
reversed precisely. The Lyapunov instability going backward in time changes sign, from attractive
to repulsive with the Lyapunov instability offset by convergence going forward but completely
uncontrolled backward despite the time-reversibility of the equations of motion.

Detailed characterizations of the fractals found in the Walk and Map
problems have revealed an unsettling nonuniformity of convergence.
Different approachs to the information dimensions of the models 
[ $\delta = (1/3)^n$ and $\delta = (1/4)^n$, for instance ]
give different results\cite{b11}!

We believe that further study of these fractals is warranted. As an
inducement the 2021 Snook Prize problem seeks to shed light on the
information dimension of the fractals, 1.7337 {\it versus}  $1.741_5$
for the Baker Map, equivalent to 0.7337 {\it versus} $0.741_5$ for
the Confined Walk problem.  Straightforward binning calculations, along
the lines of {\bf Figures 8 and 9 }, give histogram probabilities
$\{ \ ${\tt Prob}($\delta$) $\}$
which can then be analyzed for a bin-width dependent information dimension:
$$ 
D_I(\delta) = \langle \ \ln({\tt Prob}[\delta]) \ \rangle/\ln(\delta) \ .
$$ 
Computations of $D_I(\delta)$ for small bin sizes appear to lead to three
different estimates for the $\delta \rightarrow 0$ limiting ``information
dimension''! Evidently {\it the} information dimension\cite{b11,b12,b13,b14} of
these maps (in the limit that the bin-width $\delta$ vanishes) is ill-defined,
an interesting example of nonuniform convergence. For further information
on the 2021 Snook Prizes, one thousand United States dollars, see the
articles available free at CMST.eu, the web site of the open-access journal
Computational Methods in Science and Technology.

\end{document}